\documentclass[prl, amsmath,amssymb,aps, reprint, preprintnumbers, nofootinbib,superscriptaddress]
              {revtex4-2}
\usepackage{xcolor}
\usepackage{dsfont}
\usepackage{graphicx}   
\usepackage{dcolumn}    
\usepackage{bm}
\usepackage{hyperref}
\usepackage{float}
\usepackage{lipsum}
\usepackage{afterpage}
\usepackage{placeins}

\usepackage{float}

\usepackage{array}
\usepackage{comment}
\usepackage{ulem}

\usepackage{amsmath}

\def\delp{\delta \! P}

\usepackage{tikz,xcolor,hyperref}

\definecolor{lime}{HTML}{A6CE39}
\DeclareRobustCommand{\orcidicon}{\hspace{-1mm}
	\begin{tikzpicture}
	\draw[lime, fill=lime] (0,0) 
	circle [radius=0.12] 
	node[white] {{\fontfamily{qag}\selectfont \tiny \,ID}};
	\draw[white, fill=white] (-0.0525,0.095) 
	circle [radius=0.007];
	\end{tikzpicture}
	\hspace{-3mm}
}

\foreach \x in {A, ..., Z}{\expandafter\xdef\csname orcid\x\endcsname{\noexpand\href{https://orcid.org/\csname orcidauthor\x\endcsname}
		{\noexpand\orcidicon}}
}

\begin{document}

\preprint{TIFR/TH/23-2}
\title{Quantum mismatch: a powerful measure of ``quantumness" in neutrino oscillations}
\author{Dibya~S.~Chattopadhyay\orcidA{}}
\email{d.s.chattopadhyay@theory.tifr.res.in}
\affiliation{Tata Institute of Fundamental Research, Homi Bhabha Road, Colaba, Mumbai 400005, India}
\author{Amol~Dighe\orcidB{}}
\email{amol@theory.tifr.res.in}
\affiliation{Tata Institute of Fundamental Research, Homi Bhabha Road, Colaba, Mumbai 400005, India}

\date{\today}

\begin{abstract}
The quantum nature of neutrino oscillations would be reflected in the mismatch between the neutrino survival probabilities with and without an intermediate observation.
We propose this ``quantum mismatch'' as a measure of quantumness in neutrino oscillations, which precisely extracts the interference term in the two-flavor limit.
In the full three-flavor scenario, we provide modified definitions of the Leggett-Garg and quantum mismatch measures. These are applicable for long-baseline and reactor neutrino experiments that measure neutrino survival probabilities with negligible matter effects.
\end{abstract}

\maketitle

\noindent {\it Introduction} ---
Tests of quantum mechanics (QM) provide insights into the limits of local realism, which aligns with the classical worldview that all properties of physical objects have values that exist independently of their measurements.
For example, the Bell's inequality~\cite{Bell:1964kc} tests for violations of the classical upper bound on correlations between measurements made on spatially separated systems. Violations of this upper bound~\cite{Freedman:1972zza,Aspect:1982fx} clearly indicate the need for QM, 
as they would be incompatible with the hypothesis of hidden variables~\cite{Bell:1964fg,Kochen:1968zz}.

The  Leggett-Garg (LG) measure~\cite{Leggett:1985zz,Emary_2013} provides another test of ``quantumness'' (more precisely, non-classicality) of a system through the correlations between its measurements at different times.
The Leggett-Garg Inequality (LGI) tests for the \textit{interference} in QM, as opposed to \textit{entanglement}, which is tested by the Bell's inequality.
The simplest LG measure $K_3$ employs the observation of the system at an intermediate time.

In the phenomenon of neutrino oscillations, neutrinos change their flavor ($\nu_e,\, \nu_\mu,\, \nu_\tau$) during propagation due to the interference between different mass eigenstates~\cite{SajjadAthar:2021prg,ParticleDataGroup:2022pth}. This is a unique system where QM manifests itself over hundreds and thousands of kilometers, which makes it a prime candidate for tests of QM~\cite{Blasone:2007vw,Gangopadhyay:2013aha,Blasone:2014jea,Alok:2014gya,Banerjee:2015mha,Nogueira:2016qsk,Naikoo:2017fos,Song:2018bma,Naikoo:2019eec,Ming:2020nyc,Blasone:2021cau,Wang:2022tnr,Bittencourt:2022tcl,Li:2022mus}.
Violations of LGI have been measured at neutrino oscillation experiments at MINOS~\cite{Formaggio:2016cuh} and Daya-Bay~\cite{Fu:2017hky}.
New physics effects on the LG measure have been discussed in~\cite{Shafaq:2020sqo,Shafaq:2021lju}.

A difference between observations with and without an intermediate measurement would be a natural measure of quantumness~\cite{Feynman:1981tf,Sokolovski2019}.
In this Letter, we introduce the ``quantum mismatch'' measure, $\delp$, for ascertaining the quantum nature of neutrino oscillations.
It is simply defined as the difference between neutrino survival probabilities with and without an intermediate measurement. 
Here, we use measurements at different energies as proxies for measurements at different times, which ensures that the ``intermediate''  measurement is non-invasive.

In real-world neutrino experiments, it is not possible to detect all neutrino flavors. This necessitates modification of the measures $K_3$ and $\delp$ in the full three-flavor scenario. We identify the energies where the two modified measures $\widetilde K_3$ and $\widetilde{\delp}$ would be efficient in experiments.

\begin{figure}[t]
	\centering
	\begin{tikzpicture}[scale=1.5]
		\node at (1.3,3.06) {$Q(t_i):$};
		\node at (1.18,2.6) {$t_0=0$};
		\node at (1,1.8) {$t_1$};
		\node at (1,1) {$t_2$};
		
		\draw[->,thick,black] (.7,2.6) -- (.7,1) node[rotate=90,pos=0.5,yshift=.3cm] {time};
	\end{tikzpicture}
	\hspace{.15cm}
	\begin{tikzpicture}[scale=1.5]
		\filldraw[blue] (2,1) circle (2pt) node[right] {\;B};
		\filldraw[blue] (2,2.6) circle (2pt) node[right] {\;A};
		\filldraw[red] (1,1) circle (2pt) node[left] {D\;};
		\filldraw[red] (1,2.6) circle (2pt) node[left] {C\;};
		
		\draw[->, shorten >=4pt, shorten <=4pt,thick,blue] (2,2.6) -- (2,1) node[midway,right] {};
		\draw[->, shorten >=4pt, shorten <=4pt,thick,violet] (2,2.6) -- (1,1) node[midway,above] {};
		
		\node at (1,3.08) {$-1$};
		\node at (2,3.08) {$+1$};
		\node at (1,2.85) {($\nu_x$)};
		\node at (2,2.85) {($\nu_\mu$)};
	\end{tikzpicture}
	\hspace{.5cm}
	\begin{tikzpicture}[scale=1.5]
		\filldraw[blue] (2,1) circle (2pt) node[right] {\;B};
		\filldraw[blue] (2,1.8) circle (2pt) node[right] {\;X};
		\filldraw[blue] (2,2.6) circle (2pt) node[right] {\;A};
		\filldraw[red] (1,1) circle (2pt) node[left] {D\;};
		\filldraw[red] (1,1.8) circle (2pt) node[left] {Y\;};
		\filldraw[red] (1,2.6) circle (2pt) node[left] {C\;};
		
		\draw[->, shorten >=4pt, shorten <=4pt,thick,blue] (2,2.6) -- (2,1.8) node[midway,right] {$a_1$};
		\draw[->, shorten >=4pt, shorten <=4pt,thick,blue] (2,1.8) -- (2,1) node[midway,right] {$a_2$};
		\draw[->, shorten >=4pt, shorten <=4pt,thick,violet] (2,2.6) -- (1,1.8) node[midway,yshift=.36cm] {$b_1$};
		\draw[->, shorten >=4pt, shorten <=4pt,thick,red] (1,1.8) -- (1,1) node[midway,left] {};
		\draw[->, shorten >=4pt, shorten <=4pt,thick,violet] (2,1.8) -- (1,1) node[pos=0.3, above] {};
		\draw[->, shorten >=4pt, shorten <=4pt,thick,violet] (1,1.8) -- (2,1) node[pos=0.35,above] {$b_2$};
		
		\node at (1,3.08) {$-1$};
		\node at (2,3.08) {$+1$};
		\node at (1,2.85) {($\nu_x$)};
		\node at (2,2.85) {($\nu_\mu$)};
	\end{tikzpicture}
	\caption{Schematic representation of the states starting with $\nu_\mu$ at $t_0=0$ without [left] and with [right] intermediate measurements. Only two neutrino flavors, $\nu_\mu$~and $\nu_x$, are assumed.}
	\label{fig:schematic}
\end{figure}
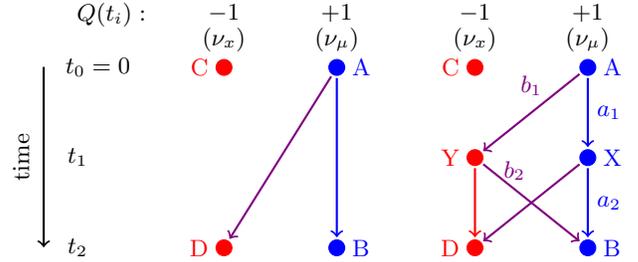

\noindent{\it Formalism and definitions} ---
In Fig.~\ref{fig:schematic}, we schematically represent the states of a system with and without an intermediate measurement.
At time $t_0=0$, the whole system is in the $\nu_\mu$ state (denoted by $\text{A}$). Over time, the neutrino flavor can either survive as $\nu_\mu$ or change to $\nu_x$. The state of the system can be measured at later times $t_1$ and $t_2$.
We denote the relevant quantum amplitudes
\begin{align}
	\mathcal{A}_{AX} \equiv \mathcal{A}_{\mu\mu} (0,t_1) =  a_1\, , \;\; & \mathcal{A}_{AY} \equiv \mathcal{A}_{ x \mu} (0,t_1) = b_1\, , \nonumber\\
	 \mathcal{A}_{XB} \equiv \mathcal{A}_{\mu\mu} (t_1,t_2) = a_2\, , \;\; & \mathcal{A}_{YB} \equiv \mathcal{A}_{ \mu x} (t_1,t_2) = b_2\, ,
\end{align}
where $\mathcal{A}_{\beta \alpha} (t_i,t_j)$ denotes the quantum amplitude for $\nu_\alpha (t_i) \to \nu_\beta (t_j)$.
The corresponding oscillation probabilities are given by $P_{\alpha \beta} (t_i, t_j) \equiv |\mathcal{A_{ \beta \alpha}} (t_i, t_j)|^2$.
Note that conservation of probability implies $|a_i|^2+|b_i|^2=1$.

In the classical limit, the muon neutrino survival probability $P_{AB} \equiv P_{\mu\mu} (0,t_2)$ would be
\begin{equation}
	P_{AB}  =P_{AX} P_{XB} + P_{AY} P_{YB} = \left| a_1 a_2  \right|^2 + \left| b_1 b_2  \right|^2 \,.
\end{equation}
In QM, in the absence of any intermediate observation, we add the amplitudes over all possible paths, obtaining
\begin{align}
	P_{AB} \equiv |\mathcal{A}_{AB}|^2
	= \left| a_1 a_2 +b_1 b_2 \right|^2\;.
\end{align}

The simplest LG measure $K_3$ is defined through a dichotomic observable $Q(t_i)$, which can only have outcomes $\pm1$.
We define $Q(t_i) = +1$ if the detected state is $\nu_\mu$, and $Q(t_i) = -1$ for any other state $\nu_x$.
The correlation function is defined as
\begin{equation}
	\mathbb{C}_{ij}\equiv \langle \, Q(t_i) \, Q(t_j) \, \rangle\;.
\end{equation}
The LG measure $K_3$ is then
\begin{equation}
	K_3 \equiv \mathbb{C}_{01} + \mathbb{C}_{12} -\mathbb{C}_{02}\;,
	\label{eq:K3}
\end{equation}
where the suffixes ($0,\, 1,\, 2$) correspond to the times ($0,\, t_1,\, t_2$).
In the classical scenario, $ -3 \le K_3 \le 1$~\cite{Leggett:1985zz,Emary_2013}, i.e., the LGI $K_3 \le 1$ is satisfied.
Any observation $K_3>1$ would indicate the quantum nature of the system.

For the muon neutrino survival probability, the quantum mismatch parameter $\delp$ is
\begin{equation}
	\delp_{\mu\mu}  = P_{AB} - \mkern-10mu \sum_{I = \text {X},\text {Y},...} \mkern-10mu P_{AI} P_{IB}\;,
\end{equation} 
where $I$ denotes the possible intermediate  neutrino flavor states ($\text{X}$,~$\text{Y}$,~...) at $t_1$.
In the classical scenario, the equality $ \delp_{\mu\mu} \!= 0$ holds. Any observation $	\delp_{\mu\mu} \! \neq 0$ would indicate the quantum nature of the system.

\smallskip
\noindent{\it Two-flavor limit} ---
in the limit of two neutrino flavors ($2\nu$~limit), conservation of probability implies $P_{x\mu}= P_{\mu x}$. The correlation function $\mathbb{C}_{ij}$ then becomes
\begin{equation}
	\mathbb{C}_{ij}^{(2\nu)} = 2\, P_{\mu\mu} (t_i , t_j)-1 \;.
\end{equation}
The LG measure $K_3$ can be calculated as
\begin{align}
	K_3^{(2\nu)}=& \, 2 \, (P_{AX} +P_{XB} - P_{AB})-1 \nonumber\\
	=&\, 1- 4 \, |b_1|^2 |b_2|^2 - 4 \, \text{Re}\left[b_1^\star b_2^\star a_1 a_2\right]\;.
\end{align}
Clearly, the quantity responsible for a possible violation of the classical bound ($K_3 \le 1$) is the interference term
\begin{equation}
	\mathcal{I}^{(2\nu)}  \equiv \text{Re}\left[b_1^\star b_2^\star a_1 a_2\right]\;.
\end{equation}
The quantum mismatch measure $\delp$ in the $2\nu$ limit is
\begin{align}
	 \delp_{\mu\mu}^{(2\nu)} 
	 = &\, P_{AB} -\! \Big( P_{AX} P_{XB} +P_{AY} P_{YB} \Big) =  2 \, \mathcal{I}^{(2\nu)},
\end{align}
which is the same interference term.
However, while $K_3^{(2\nu)}$ needs to be greater than 1 to indicate quantumness, $\delp_{\mu\mu}^{(2\nu)} \! \neq 0$ is enough to do the same.
Note that $K_3^{(2\nu)} \! >1$ necessitates $	\mathcal{I}^{(2\nu)}  \! < - |b_1|^2 |b_2|^2$, whereas for all $\mathcal{I}^{(2\nu)} \! \neq 0$, we obtain $\delp_{\mu\mu}^{(2\nu)} \! \neq 0$.
\smallskip

\noindent{\it Two-flavor quantum measures at neutrino oscillation experiments} ---
The quantum measures discussed above need measurements of the system corresponding to three different time intervals $\Delta t_{10} \equiv t_1 -t_0$, $\Delta t_{21} \equiv t_2 -t_1$ and $\Delta t_{20} \equiv t_2 -t_0$. In a fixed-baseline neutrino experiment, measurements at multiple time intervals are not possible. However, this obstacle may be overcome as follows.

Consider the evolution of a particle with mass $m$ during time interval $\Delta \tau$ in its rest frame. If this particle is observed to have an energy $E$ in the lab frame, the same evolution will be observed for a time interval $\Delta t = (\Delta \tau /m)\, E$, by the special theory of relativity. Thus, the evolution of a neutrino in the lab frame depends only on the ratio $\Delta t/ E$.

For neutrino oscillations, this dependence on $\Delta t/ E$ holds in vacuum, or as long as matter effects~\cite{Wolfenstein:1977ue,Mikheyev:1985zog} are negligible.
In this limit, the measurements of neutrinos with the same energy at different times may be replaced by measurements of neutrinos of different energies at the same time intervals. That is, for some energy $E_0$ and time interval $\Delta t_0$, if we find
\begin{equation}
	\left( \frac{\Delta t_{10}}{E_0},\; \frac{\Delta t_{21}}{E_0},\; \frac{\Delta t_{20}}{E_0} \right) = \left( \frac{\Delta t_{0}}{E_{10}},\; \frac{\Delta t_{0}}{E_{21}},\; \frac{\Delta t_{0}}{E_{20}} \right)\;,
\end{equation}
then the measurements at energies $E_{10}$, $E_{21}$ and $E_{20}$ can act as the proxies for measurements with time intervals $\Delta t_{10}$, $\Delta t_{21}$ and $\Delta t_{20}$, respectively. Here, $\Delta t_0$ should be taken as the duration of neutrino propagation from the source to the detector.
Since the three time intervals should be related to each other as
\begin{equation}
	\Delta t_{10}+\Delta t_{21} =\Delta t_{20}\;,
\end{equation}
the proxy energies need to satisfy the relation
\begin{equation}
	 1/E_{10} +1/E_{21} =1/E_{20} \;.
	 \label{eq:harmonicsum}
\end{equation}
In principle, for every value of $E_0$, one has a different triplet ($E_{10}$, $E_{21}$, $E_{20}$), using which the quantum measures may be defined.
Note that, using measurements at different energies makes this a truly non-invasive measurement, which does not disrupt the system in any way.

\begin{figure}[t]
	\centering
	\vspace{-.2cm}
	\includegraphics[width=0.48\linewidth]{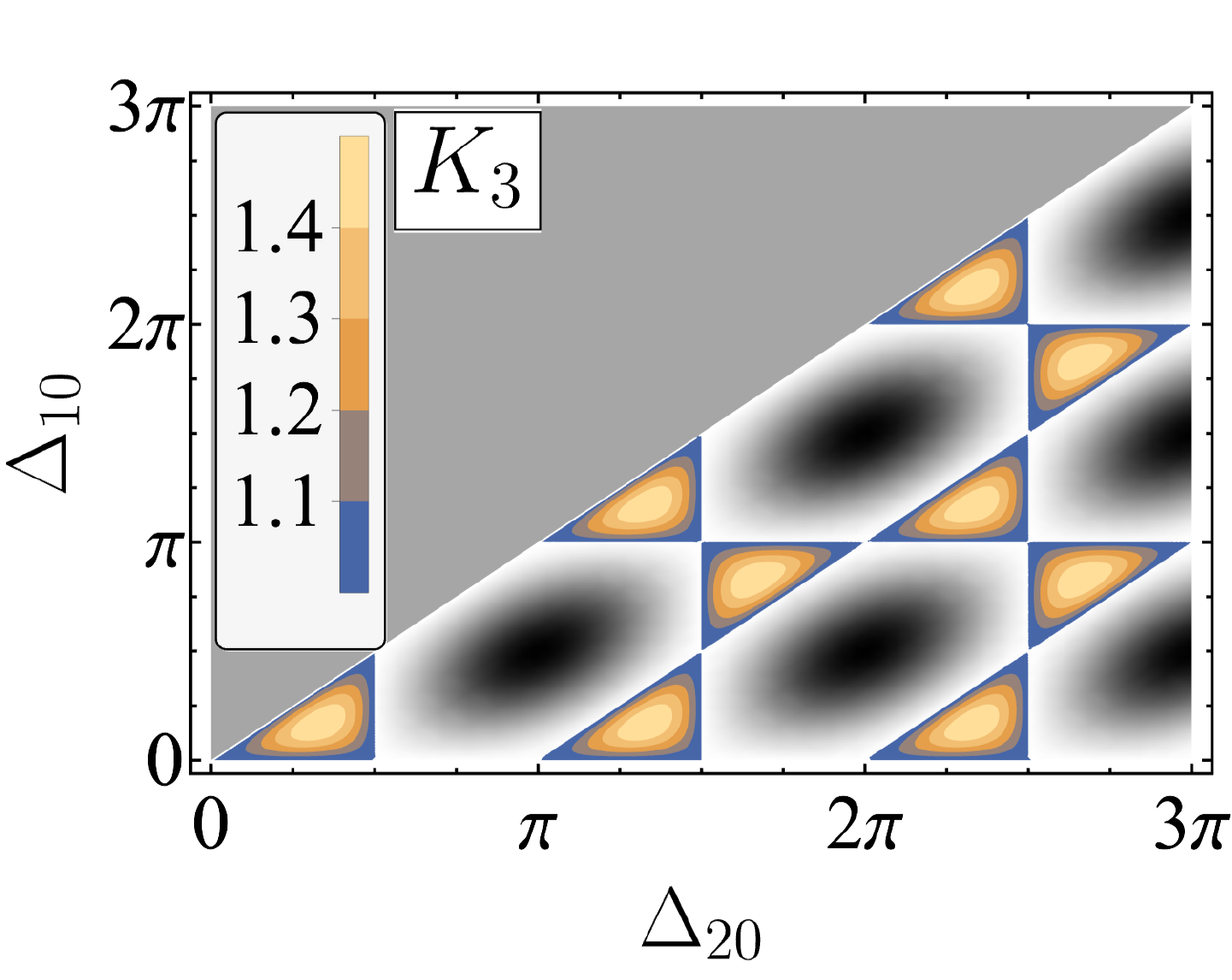}
	\hspace{.05cm}
	\includegraphics[width=0.48\linewidth]{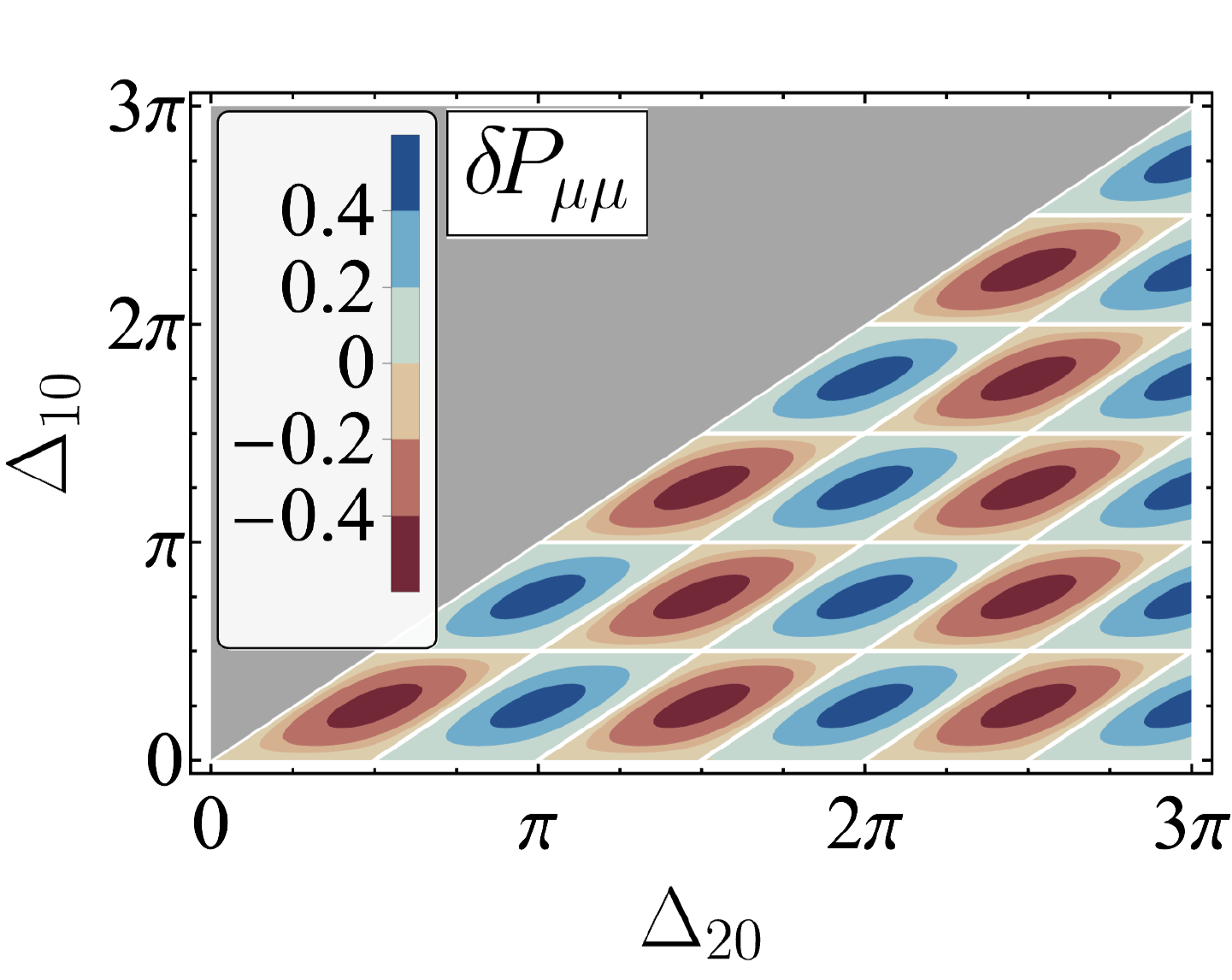}\\
	\vspace{-.05cm}
	\includegraphics[width=0.465\linewidth]{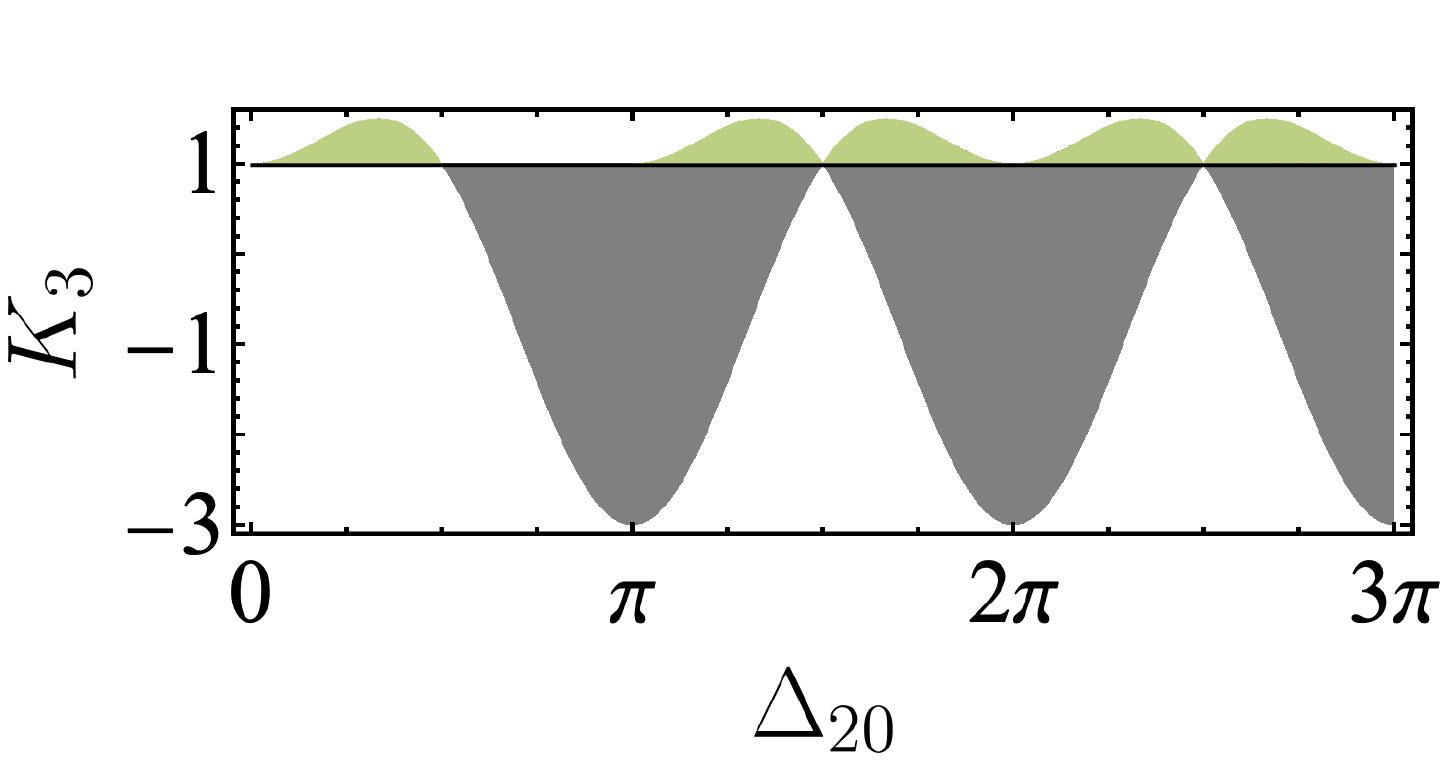}
	\hspace{.08cm}
	\includegraphics[width=0.49\linewidth]{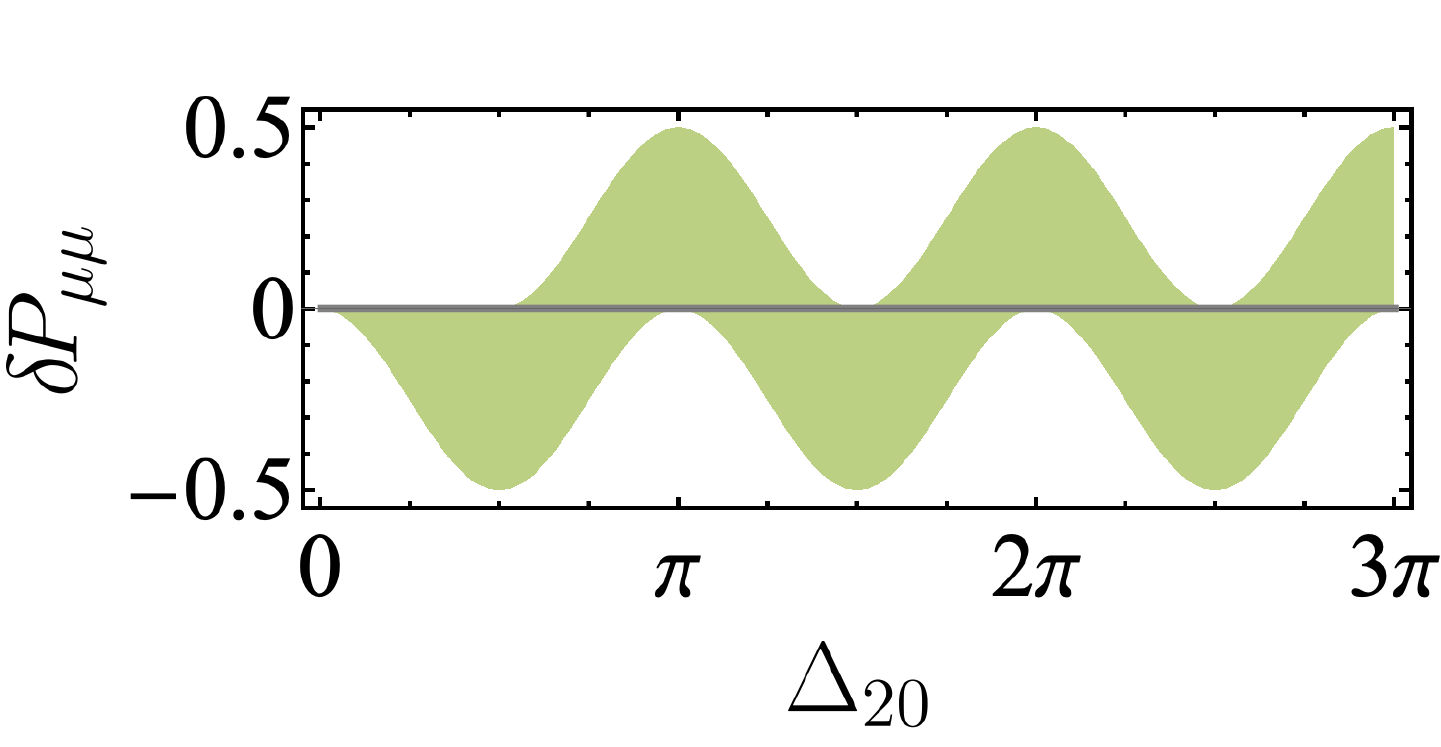}
	\caption{Top Panel: the quantum measures $K_3$ and $\delp_{\mu\mu}$ in the two-flavor limit, in the $(\Delta_{20},\, \Delta_{10})$ plane. We have taken the mixing angle $\theta=45^\circ$. The upper-left solid gray triangles are unphysical regions. The black and white regions obey the classical limit. Colored regions correspond to $K_3 >1$ and $\delp_{\mu\mu} \neq 0$, indicating quantumness. Bottom Panel: ranges of $K_3$ and $\delp_{\mu\mu} $ as functions of $\Delta_{20}$, for all possible $\Delta_{10}$. The dark gray (light green) regions obey (violate) classical limits.}
	\label{fig:2nu}
\end{figure}

The neutrino survival probability in the $2\nu$ limit is given by
\begin{equation}
	P_{\mu\mu} (t_i,t_j) = 1- \sin^2 (2\theta)\, \sin^2 \Delta\;,
	\label{eq:Pmumu}
\end{equation}
where $\theta$  is the mixing angle. The oscillation phase, as a function of energy, is
\begin{equation}
	\Delta (E) =1.27 \times (\Delta m^2\text{ in eV}^2) \times \frac{(L\text{ in km})}{(E \text{ in GeV})}\;,
\end{equation}
where $\Delta m^2$ is the mass-squared difference between the two neutrinos, and $L= c \,(t_j -t_i) $. Note that the dependence on the ratio $\Delta t/E$ is explicitly present above.

The top panels of Fig.~\ref{fig:2nu} show the values of the quantum measures $K_3$ and $\delp_{\mu\mu}$ in the $2\nu$ limit, in the $(\Delta_{20}, \,\Delta_{10})$ plane where $\Delta_{ij} \equiv \Delta (E_{ij})$. Since $\Delta_{20}$ corresponds to the largest time interval, we need $\Delta_{20} > \Delta_{10}$, making the upper-left triangles in the contour plots unphysical.
The figure shows that the classical bound of $K_3  \le 1$ would be violated in certain colored `islands' in the parameter space, whereas the classical value of $\delp_{\mu\mu} =0$ would be violated for much larger regions of possible $(\Delta_{20},\, \Delta_{10})$ choices.

The bottom panels of Fig.~\ref{fig:2nu} further illustrate that in the $2\nu$ limit, the quantum mismatch measure $\delp$ would be a more efficient probe of non-classicality.
\smallskip

\noindent {\it Defining the three-flavor quantum measures} --- 
The above discussion implicitly assumes that there are only two neutrino flavors.
However, as neutrinos come in three flavors $(3\nu)$, these measures will have to be modified accordingly.

\begin{figure}[t]
	\centering
	\begin{tikzpicture}[scale=1.6]
		\node at (1.3,3.18) {$Q(t_i):$};
		\node at (1.18,2.6) {$t_0=0$};
		\node at (1,1.8) {$t_1$};
		\node at (1,1) {$t_2$};
		
		\draw[->,thick,black] (.7,2.6) -- (.7,1) node[rotate=90,pos=0.5,yshift=.3cm] {time};
	\end{tikzpicture}
	\hspace{.6cm}
	\begin{tikzpicture}[scale=1.6]
		\filldraw[blue] (3,1) circle (2pt) node[right] {\;B};
		\filldraw[blue] (3,1.8) circle (2pt) node[right] {\;X};
		\filldraw[blue] (3,2.6) circle (2pt) node[right] {\;A};
		\filldraw[red] (2,1) circle (2pt) node[left] {D\;};
		\filldraw[red] (2,1.8) circle (2pt) node[left] {Y\;};
		\filldraw[red] (2,2.6) circle (2pt) node[left] {C\;};
		\filldraw[red!80!blue,opacity=1] (1,1) circle (2pt) node[left] {F\;};
		\filldraw[red!80!blue,opacity=1] (1,1.8) circle (2pt) node[left] {Z\;};
		\filldraw[red!80!blue,opacity=1] (1,2.6) circle (2pt) node[left] {E\;};
		
		\draw[->, shorten >=4pt, shorten <=4pt,thick,blue] (3,2.6) -- (3,1.8) node[midway,right] {$a_1$};
		\draw[->, shorten >=4pt, shorten <=4pt,thick,blue] (3,1.8) -- (3,1) node[midway,right] {$a_2$};
		\draw[->, shorten >=4pt, shorten <=4pt,thick,violet] (3,2.6) -- (2,1.8) node[xshift=.75cm,yshift=.3cm] {$b_1$};
		\draw[->, shorten >=4pt, shorten <=4pt,thick,violet] (2,1.8) -- (3,1) node[pos=0.5,above] {$b_2$};
		\draw[->, shorten >=4pt, shorten <=4pt,thick,violet] (3,2.6) -- (1,1.8) node[pos=0.7,yshift=.22cm] {$c_1$};
		\draw[->, shorten >=4pt, shorten <=4pt,thick,violet] (1,1.8) -- (3,1) node[pos=0.45,below] {$c_2$};
		
		\node[draw,fill=red!70,opacity=0.3,minimum width=2.45cm,minimum height=.5cm,anchor=center] at (1.45,1.8) {};
		\node[draw,fill=red!70,opacity=0.3,minimum width=2.45cm,minimum height=.5cm,anchor=center] at (1.45,1) {};
		\node[draw,fill=red!70,opacity=0.3,minimum width=2.45cm,minimum height=.5cm,anchor=center] at (1.45,2.6) {};
		
		\node at (1,3.18) {$-1$};
		\node at (2,3.18) {$-1$};
		\node at (3,3.18) {$+1$};
		\node at (1,2.95) {($\nu_\tau$)};
		\node at (2,2.95) {($\nu_e$)};
		\node at (3,2.95) {($\nu_\mu$)};
	\end{tikzpicture}
	\caption{Schematic representation of the states of the system with three neutrino flavors, starting with $\nu_\mu$ at $t_0=0$.}
	\label{fig:3nuschematic}
\end{figure}
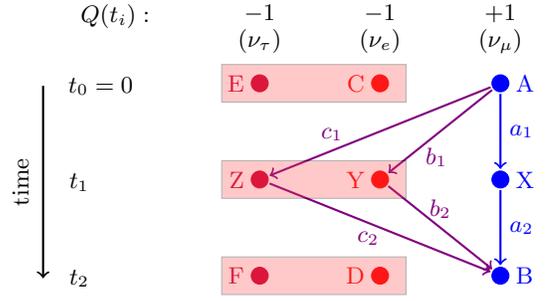

Since a dichotomic observable $Q(t_i)$ is needed for the LG measure, we shall assign $Q (t_i) =-1$ for all non-muon neutrinos, i.e. $\nu_e$ and $\nu_\tau$, as depicted in Fig.~\ref{fig:3nuschematic}.

If we had the ability to detect all three neutrino flavors, i.e., if independent measurements of neutrino flavor states $X$, $Y$ and $Z$ were possible, the correlation functions $\mathbb{C}_{01}$ and $ \mathbb{C}_{02}$ would take the same form as before:
\begin{equation}
	\mathbb{C}_{01}^{(3\nu)} = 2 \, P_{AX}-1\, , \quad \mathbb{C}_{02}^{(3\nu)}  = 2\, P_{AB} -1\,.
\end{equation}
The correlation function $\mathbb{C}_{12}$, however, would be
\begin{align}
	\mathbb{C}_{12}^{(3\nu)} \! = &\, P_{AX} \left( P_{XB} - P_{XD} - P_{XF}  \right) \nonumber\\
	& - \sum_{I =Y, Z} P_{AI} \left( P_{IB} - P_{ID} - P_{IF}  \right) \\
	= & 1 + 2 \, (P_{AX} P_{XB} - \! P_{AY} P_{YB} -  \! P_{AZ} P_{ZB})- \! 2 \, P_{AX} \nonumber \,.
\end{align}
The LG measure $K_3$ can then be calculated using Eq.~(\ref{eq:K3}).

In QM, in the absence of any intermediate observation,
\begin{equation}
	P_{AB}= |a_1 a_2 + b_1 b_2 + c_1 c_2|^2\;.
\end{equation}
The value of $K_3$ would then be
\begin{align}
	K_3^{(3\nu)} =&\, 1 - 4\, |b_1|^2 |b_2|^2 -4\, |c_1|^2 |c_2|^2 - 4 \, \mathcal{I}^{(3\nu)},
\end{align}
where we have defined the $3\nu$-interference term as,
\begin{equation}
	\mathcal{I}^{(3\nu)}  \equiv \text{Re} [a_1^\star a_2^\star b_1 b_2] + \text{Re} [b_1^\star b_2^\star c_1 c_2] + \text{Re} [c_1^\star c_2^\star a_1 a_2]\,.
\end{equation}
Clearly, in the absence of the interference terms, the classical bound of $K_3^{(3\nu)} \! \le 1$ will be always satisfied.

For experiments where all neutrino flavors cannot be detected, $K_3^{(3\nu)}$ as defined above cannot be measured. However, a modified LG measure, observable at all experiments which can measure the muon neutrino survival probability $P_{\mu\mu}$, can be defined as
\begin{equation}
	\widetilde{K}_3 = \widetilde {\mathbb{C}}_{01} +  \widetilde {\mathbb{C}}_{12} - \widetilde {\mathbb{C}}_{02}\;,
\end{equation}
where $\widetilde {\mathbb{C}}_{ij} $ is defined as
\begin{equation}
	\widetilde {\mathbb{C}}_{ij} = 2\, P_{\mu\mu} (t_i, \, t_j)-1\;.
\end{equation}
This modified definition of LG measure makes it directly applicable for those long-baseline neutrino experiments where matter effects are negligible for $P_{\mu\mu}$.
We get
\begin{align}
	\widetilde{K}_3 = & \; 2\,  \Big( P_{AX} + P_{XB} - P_{AB} \Big) -1 \nonumber\\
	=& \;  1 - 4 \, |b_1|^2 |b_2|^2 -4\,  |c_1|^2 |c_2|^2 \nonumber\\
	& - 2\,  |b_1|^2 |c_2|^2 - 2\,  |b_2|^2 |c_1|^2 - 4 \, \mathcal{I}^{(3\nu)} \;.
\end{align} 
The $3\nu$-interference term allows the violation of the classical bound $\widetilde{K}_3 \le 1$, albeit for a smaller region of parameter space compared to $K_3^{(3\nu)} $.
Thus $\widetilde{K}_3$ is a practical LG measure in the three-flavor system of neutrinos. It is indeed the one implicitly being used in~\cite{Formaggio:2016cuh,Fu:2017hky}.

Similarly, if we had the ability to detect all three neutrino flavors, the quantum mismatch measure would be
\begin{align}
	\delp_{\mu\mu}^{(3\nu)} \equiv & \, P_{AB} - \! \left( P_{AX}P_{XB} + P_{AY}P_{YB} + P_{AZ}P_{ZB} \right) \nonumber\\
	= & \, 2 \, \mathcal{I}^{(3\nu)}\;.
\end{align}
Though this measure precisely extracts the interference term that causes violations of classical bounds, it cannot be calculated for real-world experiments.
For experiments which only observe $P_{\mu\mu}$, we define the modified quantum mismatch measure
\begin{align}
	\widetilde{\delp}_{\mu\mu} \equiv & \; P_{AB} -\Big( P_{AX} P_{XB} + \left(1-P_{AX} \right) \left(1-P_{XB} \right) \Big)\nonumber \\
	= & \; \mathcal{I}^{(3\nu)} - \Big(|b_1|^2 |c_2|^2 + |b_2|^2|c_1|^2\Big)\;,
\end{align}
where we have used $P_{XB} + P_{YB} + P_{ZB} =1$, which is true due to probability conservation.
In the classical limit, i.e., in absence of the $3\nu$-interference term, $\widetilde{\delp}_{\mu\mu} \! \le 0$.
Therefore $\widetilde{\delp}_{\mu\mu}  >0$ is a clear indicator of quantumness.

Note that the modified measure $\widetilde{\delp}_{\mu\mu} $ has a classical upper bound as opposed to the ${\delp}^{(3\nu)}_{\mu\mu} $ which would have had a fixed value of zero in the classical limit. However, $\widetilde{\delp}_{\mu\mu} $ is a practical measure that can be determined at real-world experiments.

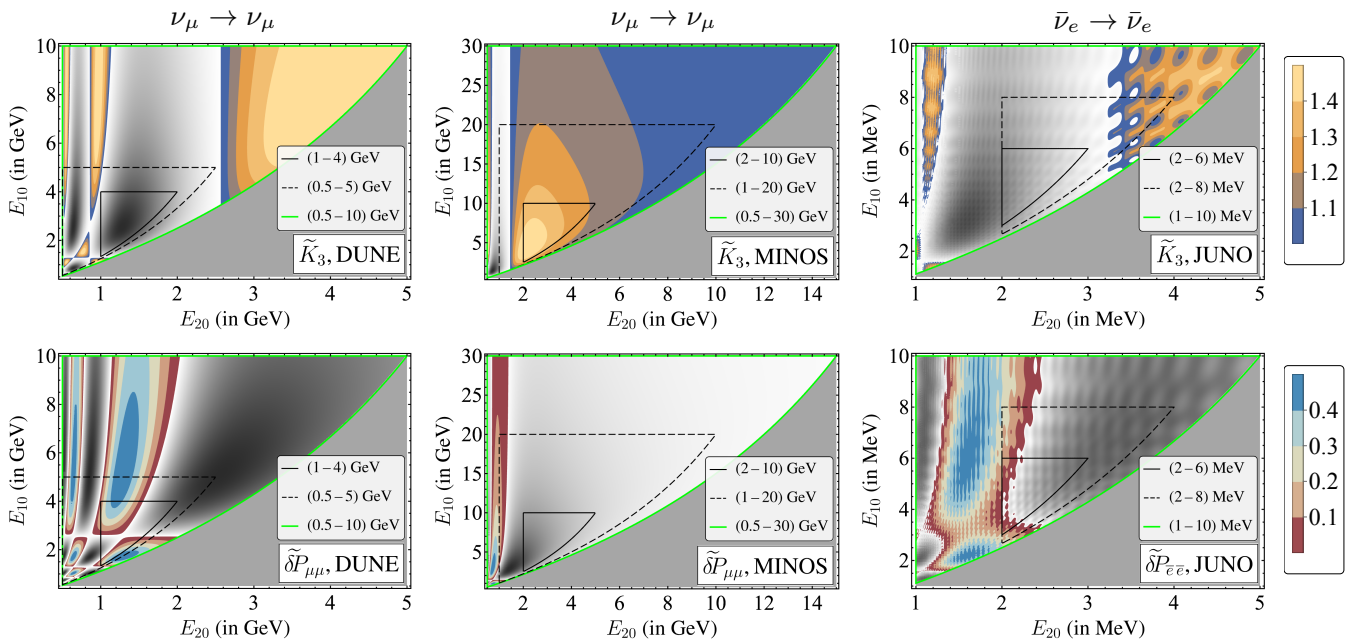
\begin{figure*}[t]
	\centering
	\begin{tikzpicture}[scale=1.5]
		\node at (0,1) {\scalebox{1.2}{$\nu_\mu \to \nu_\mu$}};
		\node at (3.9,1) {\scalebox{1.2}{$\nu_\mu \to \nu_\mu$}};
		\node at (7.8,1) {\scalebox{1.2}{$\bar \nu_e \to \bar \nu_e$}};
		\node at (8.5,1) {};
	\end{tikzpicture}\\
	\vspace{-.1cm}
	\includegraphics[width=0.3\linewidth]{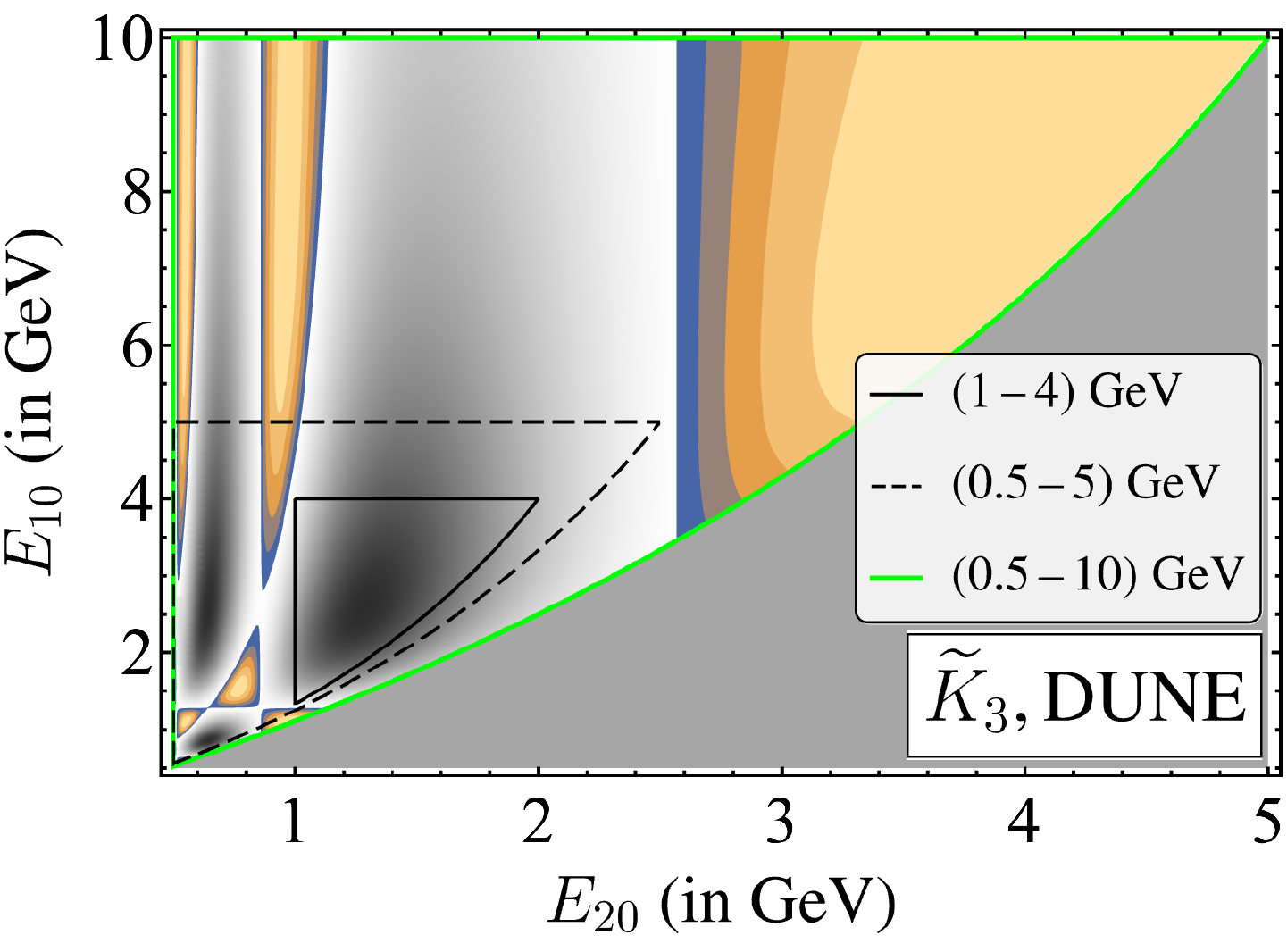}
	\hspace{.08cm}
	\includegraphics[width=0.3\linewidth]{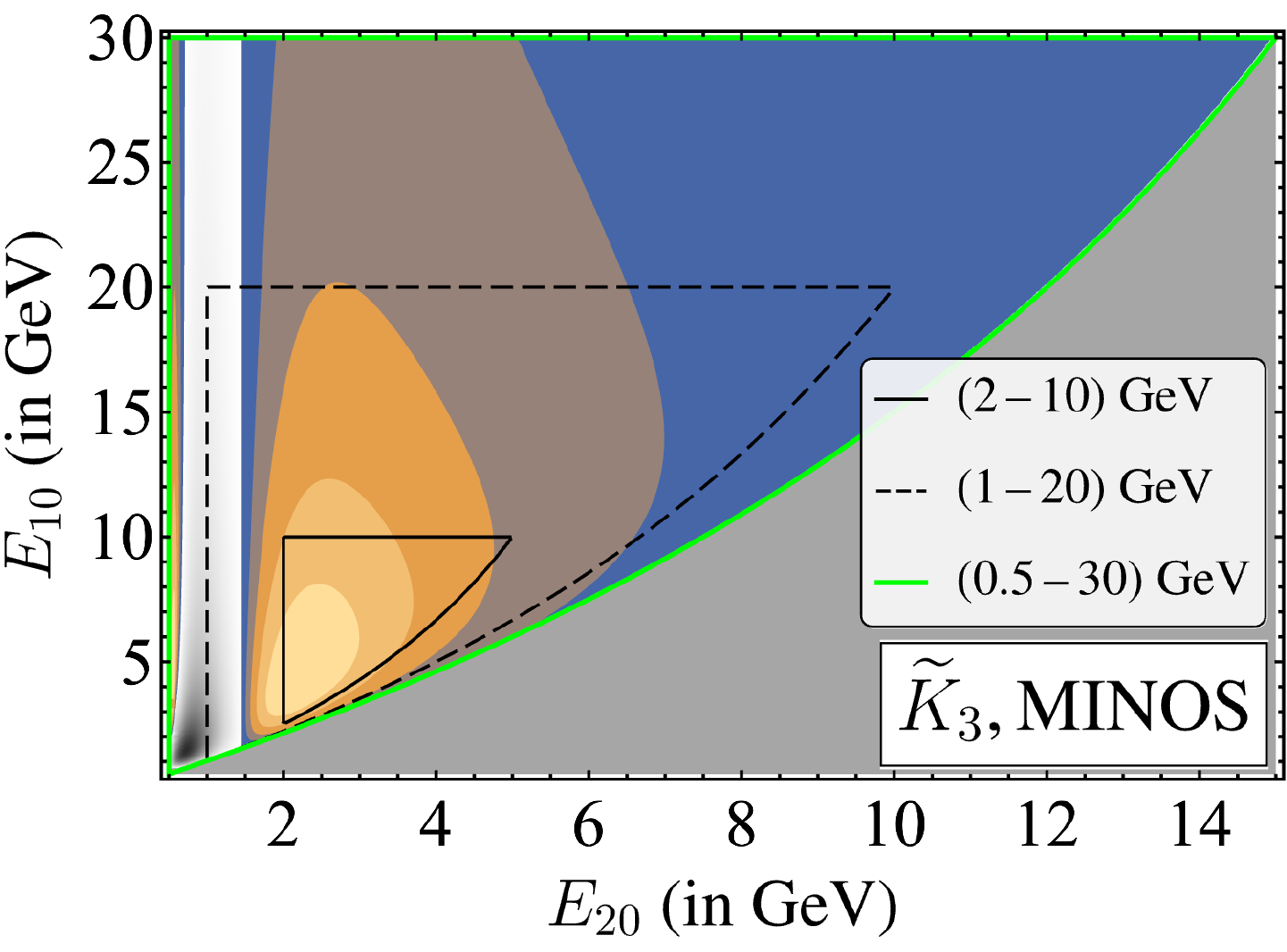}
	\hspace{.08cm}
	\includegraphics[width=0.3\linewidth]{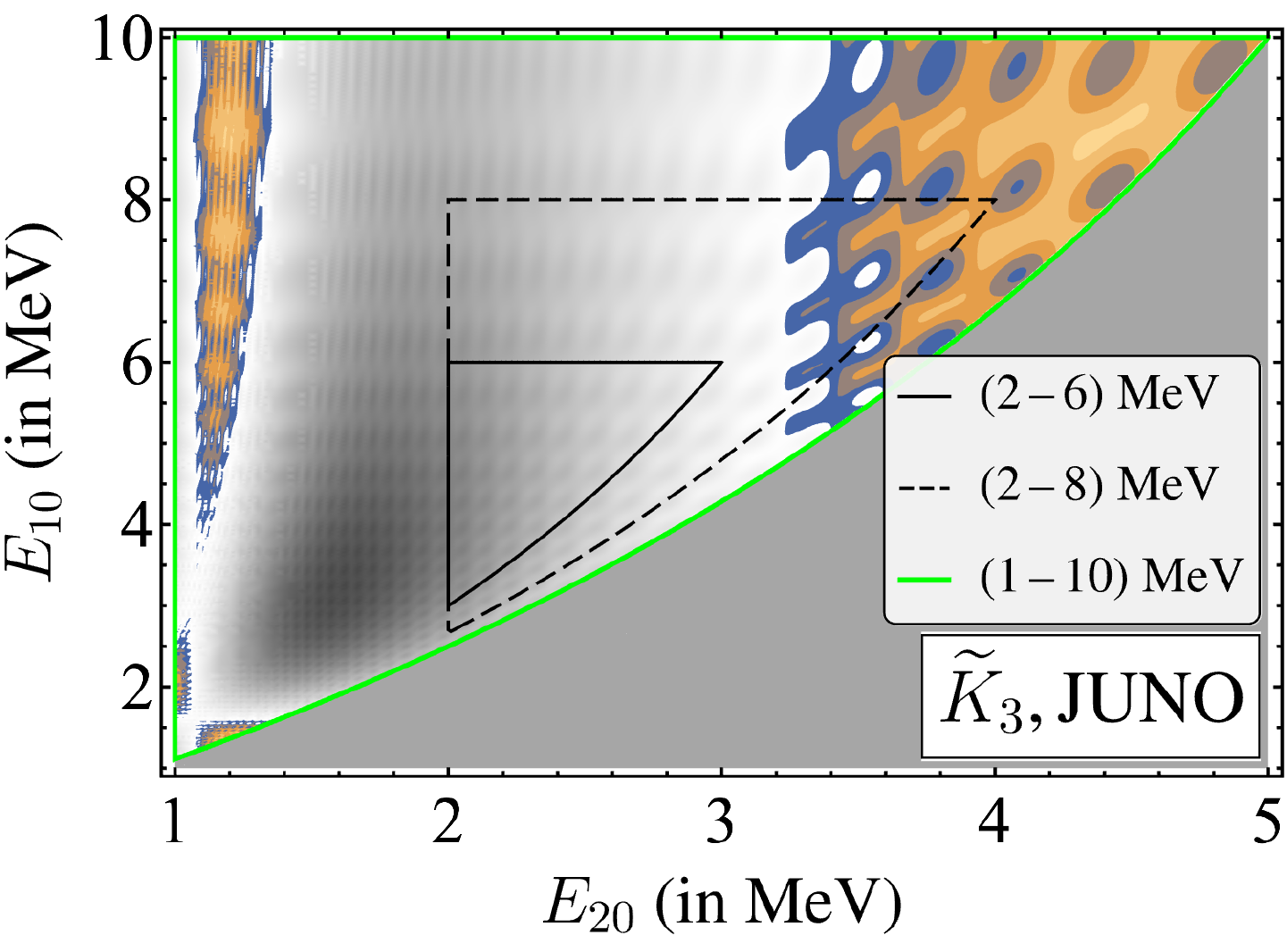}
	\hspace{.02cm}
	\raisebox{.9cm}{\includegraphics[width=0.045\linewidth]{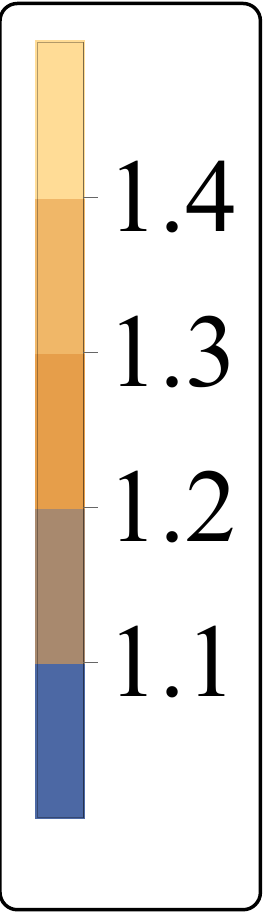}}\\
	\vspace{.1cm}
	\includegraphics[width=0.3\linewidth]{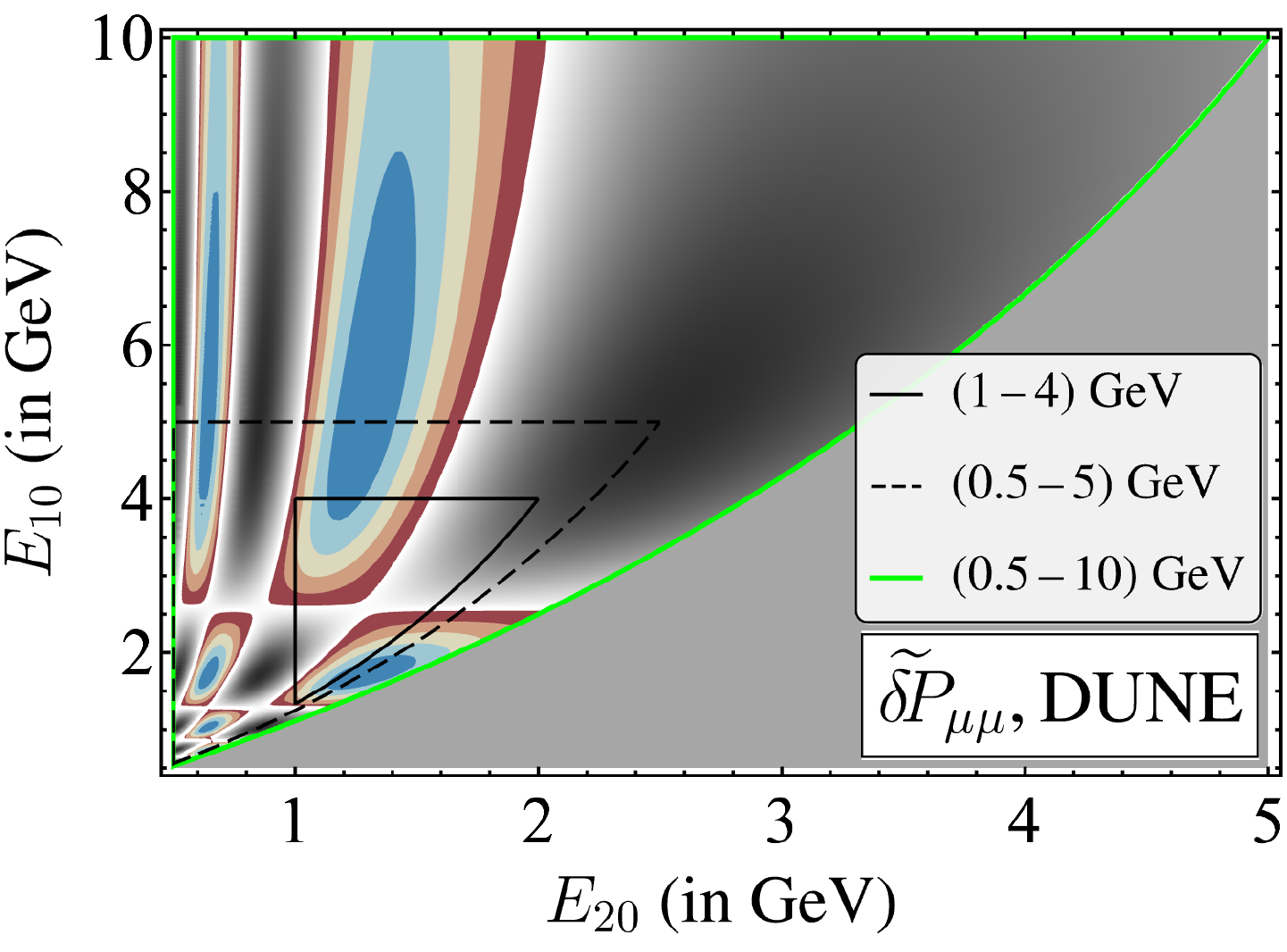}
	\hspace{.08cm}
	\includegraphics[width=0.3\linewidth]{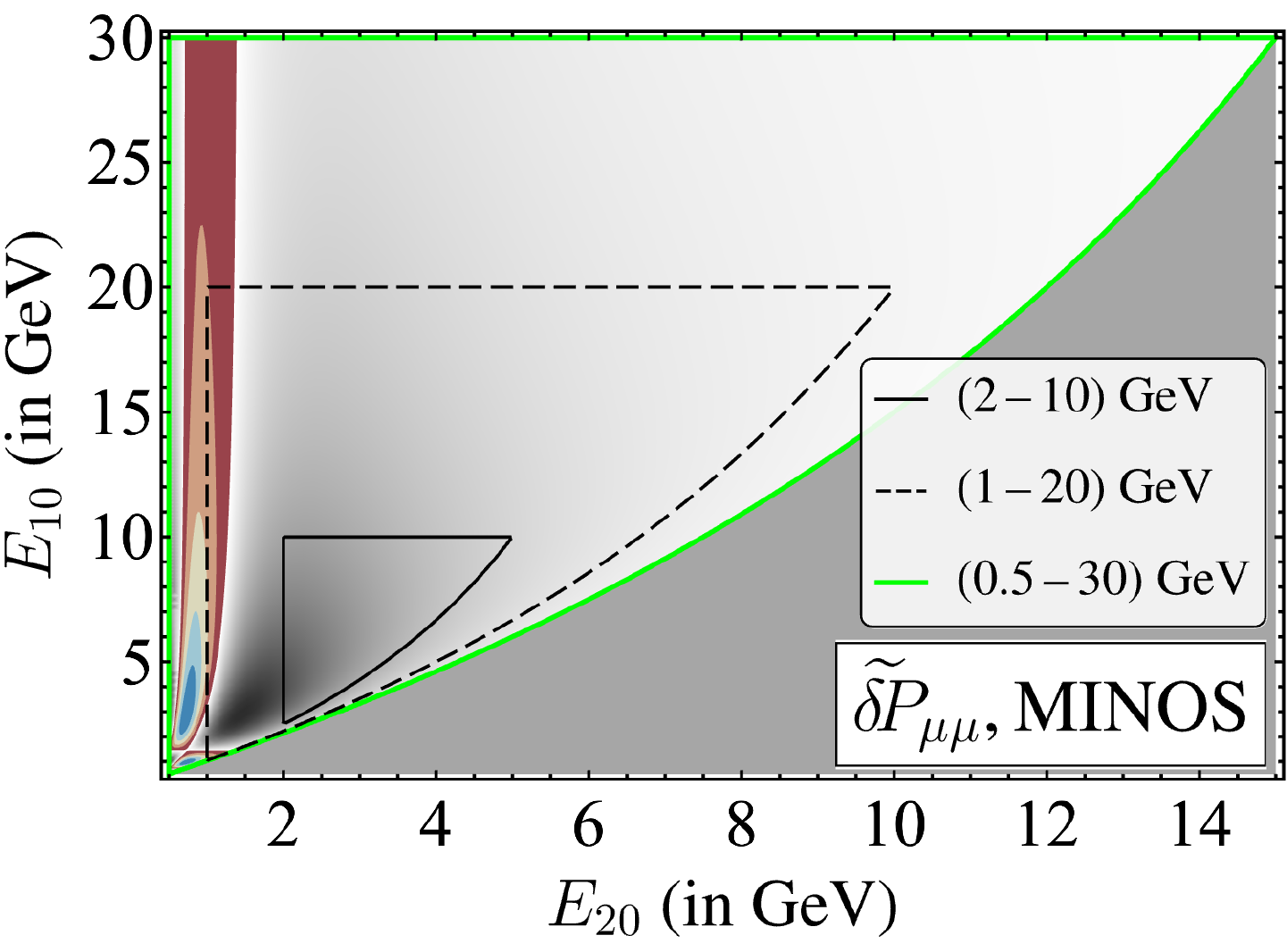}
	\hspace{.08cm}
	\includegraphics[width=0.3\linewidth]{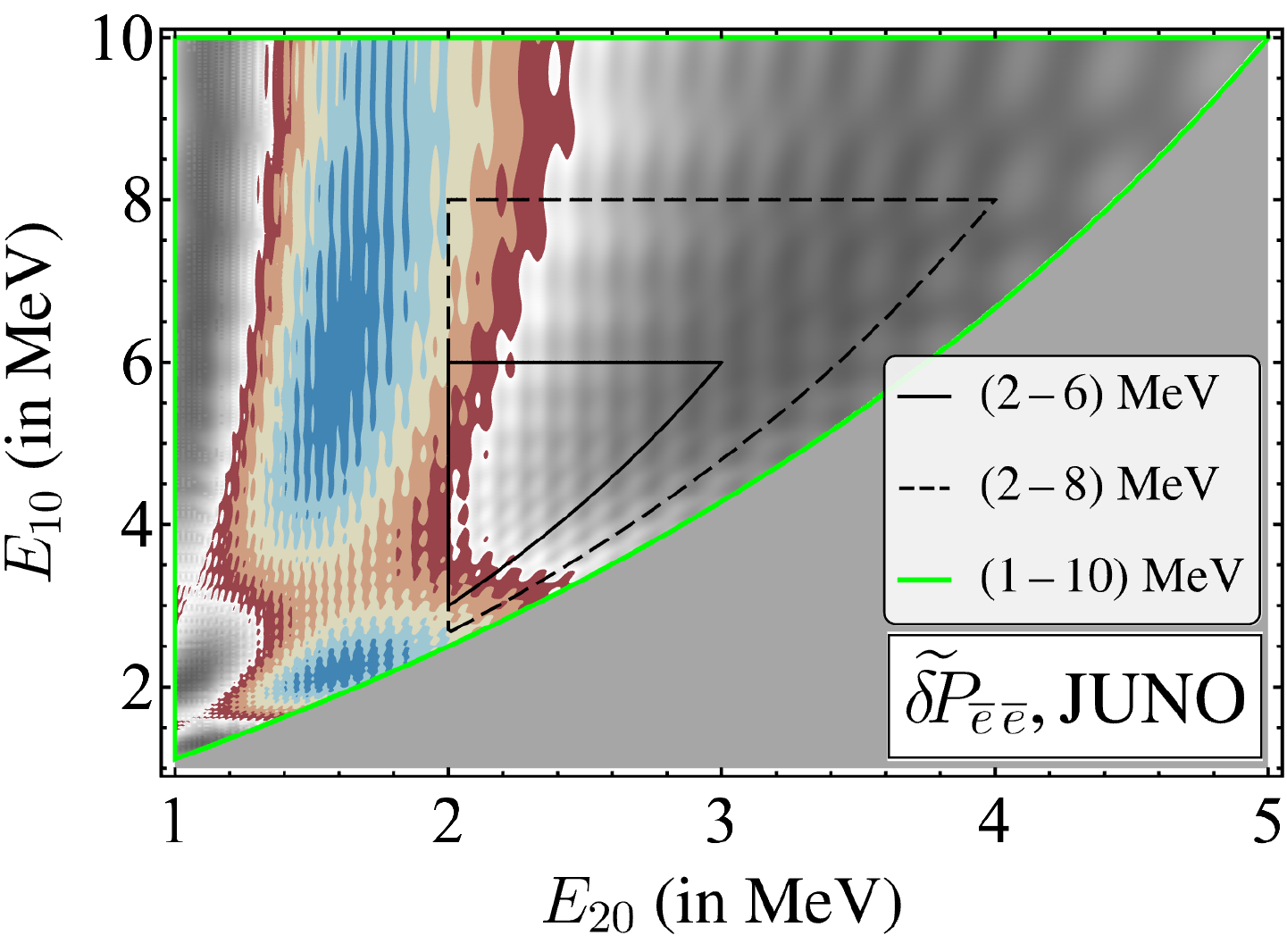}
	\hspace{.02cm}
	\raisebox{.9cm}{\includegraphics[width=0.045\linewidth]{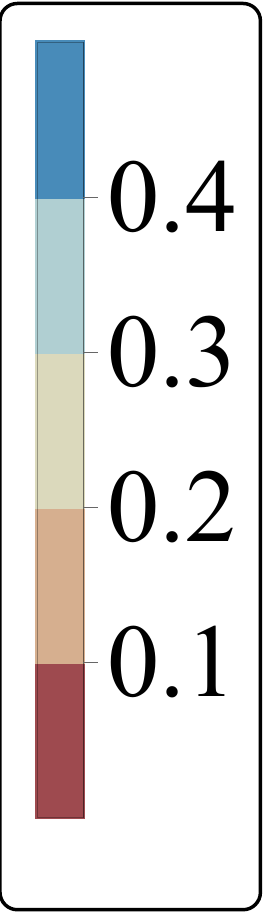}}
	\caption{
		The modified quantum measures $\widetilde{K}_3$ [top] and $\widetilde{\delp}$ [bottom] at DUNE [left], MINOS [middle] and JUNO~[right] in the the $(E_{20},\,E_{10})$ plane.
		The solid gray regions outside the green boundaries do not give an allowed $E_{21}$ value.
		The quasi-triangular regions enclosed by dashed/ solid black boundaries denote energies where the flux is higher.
		The black and white regions obey the classical limit. Colored regions correspond to $\widetilde{K}_3 >1$ and $\widetilde{\delp}> 0$, indicating quantumness.
		Neutrino parameters are given in Eq.~(\ref{eq:nuparameters}).
		The spotted features for JUNO are due to the co-existence of atmospheric and solar neutrino oscillations.}
	\label{fig:3nu}
\end{figure*}

\noindent{\it Quantum measures at neutrino oscillation experiments} ---
Although the phenomenon of neutrino oscillation is inherently quantum,
the observability of quantumness depends on the quantum measure employed as well as the parameters of the experiment. Here, we identify the energies at which the quantum nature would be observable through $\widetilde{K}_3$ and $\widetilde{\delp}$ at neutrino oscillation experiments.

In Fig.~\ref{fig:3nu}, we show the values of the modified quantum measures $\widetilde{K}_3$ and $\widetilde{\delp}$ in terms of the energies $(E_{20}, \, E_{10})$.
The value of $E_{21}$ can be obtained from Eq.~(\ref{eq:harmonicsum}).
Since $E_{20} < E_{10}$, and since all three energies must lie within the energy range of the experiment, the solid gray regions in lower-right corners of all panels are not allowed.

For the purpose of illustration, we choose the experiments DUNE and MINOS which measure $P_{\mu\mu} \equiv P (\nu_\mu \to \nu_\mu)$.
This is valid because $P_{\mu\mu}$ does not have any leading-order matter contributions~\cite{Akhmedov:2004ny}.
We further analyze the modified quantum measures through $P_{\bar e \bar e} \equiv P (\bar \nu_e \to \bar\nu_e)$ for the reactor antineutrino experiment JUNO, where matter effects are negligible.
We take neutrino parameter values consistent with the global fits~\cite{deSalas:2020pgw,Esteban:2020cvm,Capozzi:2021fjo,nufit22}:
\begin{align}
	\Delta m^2_{21} = 7.5 \times 10^{-5} \text{ eV}^2, & \quad \Delta m^2_{31} =2.5 \times 10^{-3} \text{ eV}^2, \nonumber\\
	\theta_{13} = 8.5^\circ, \; \theta_{23}= 45^\circ, \; & \theta_{12}=33.5^\circ, \; \delta_{\text{CP}} = -90^\circ  .
	\label{eq:nuparameters}
\end{align}

We observe that for DUNE~\cite{DUNE:2020jqi}, especially in the region where we expect the maximum neutrino flux, the measure $ \widetilde{\delp} \! > \!0$ is much better-suited for probing quantumness than $\widetilde{K}_3$. The results obtained for T2K/ T2HK~\cite{T2K:2011qtm,Hyper-KamiokandeProto-:2015xww} are also quite similar to this.
On the other hand, for MINOS~\cite{MINOS:2014rjg}, the modified LG measure $\widetilde{K}_3$
is more efficient. Note that the probe of quantum nature of neutrino oscillations at MINOS~\cite{Formaggio:2016cuh} implicitly uses this modified measure. We find that in NOvA~\cite{NOvA:2019cyt}, neither of the measures would be efficient in probing the quantum nature. 

For JUNO~\cite{JUNO:2015zny}, $ \widetilde{\delp}$ is a better measure in the energy range with higher flux. The energy resolution of the detector will play a crucial role in determining the observability of quantumness.

\noindent {\it Concluding remarks}  ---
Tests of violation of local realism are instrumental for probing the non-classical nature of physical systems.
In this Letter, we introduce the quantum mismatch measure $\delp$ for detecting quantumness in neutrino oscillations.
In the two-flavor limit, this measure precisely extracts the quantum interference term.

We extend the definitions of $\delp$ and the Leggett-Garg measure $K_3$ to the full three-flavor scenario.
In the absence of experiments which can detect all three neutrino flavors separately, we provide modified practical definitions of both the measures, $\widetilde{\delp}$ and $\widetilde{K}_3$, that employ only the neutrino survival probabilities.
We further identify the energy ranges at neutrino experiments where the quantum nature would become detectable.

The new quantum mismatch measure $\widetilde{\delp}$ is thus a robust, practical and efficient measure, which would further advance the quest of probing quantumness at macroscopic length scales.

\smallskip
\noindent {\it Acknowledgments} ---
We acknowledge support from the Department of Atomic Energy (DAE),
Government of India, under Project Identification No. RTI4002. We would also like to thank K.~Damle, B.~Dasgupta, P.~Mehta, R.~Sensarma and V.~Tripathi for discussions and useful comments.


\end{document}